\documentstyle[a4]{article}
\def\beq{\begin{equation}}
\def\eeq{\end{equation}}

\begin{document}
\title{Some conjectures looking for a NCG theory}

\author{Alejandro Rivero \thanks {\tt rivero@sol.unizar.es} \\
{\it Dep. F\'{\i}sica Te\'orica, Universidad de Zaragoza,
                                                50009 Zaragoza, Spain}}

\maketitle
\begin{abstract}
It is pointed out that ambiguities in the regularization of
actions with second derivatives seem to happen with the same 
multiplicity that the standard model of elementary particles
\end{abstract}

\section{Introduction, apology and apologetics}

First of all, I must apologize by bringing up a such speculative
paper as this is. Our justification is to open a forgot question
which could fit in the current research for the fine structure
of differential geometry \cite{connes}. While the anticommutativity of
Cartan differentials has the obvious appeal for physicists, the 
traditionally troubled
development of geometry has precluded us of getting
to this point at the adequate moment:
Cartan
theory itself was solidified lot of time before we got to zero the research in
a set of four elemental particles and some multiplicities and generations.
And even if this "tetrad" of particles could have constituted a clue, we 
have waited another quarter of century to get a theory of deformed, non
commutative, differential calculus able to contain it \cite{connes.cham}. 

Cartan differentials are fundamental geometrical objects, they
carry the modern interpretation of Poincare principle, a prerequisite
to get the fundamental theorem of integration. A future theory based
on such object would be epistemologically as fundamental as the
curvature-based relativity theory. But it would be probably a
fool attempt to try do derive such theory directly from
the sky of mathematical principles. 
Instead of this, we must to
try to stick to earth; our model should be as near to possible
to the observed spectrum. In fact our initial
inspiration comes from the more pragmatic approach to field
theory, namely lattice QFT.  

Lattice theorists usually find themselves fighting against unwanted
degrees of freedom coming from (supposed) discretization artifact:
fermion doubling, Gribov copies, FP doublers and so on. Such
infection is supposed to disappear in the continuum limit. But on
the other side, arguments as triviality, localization, gravity etc
drive to consider the existence
of a true natural cutoff, a minimum lattice length. It seems difficult to
marry both views.

In this paper we stop confronting head-to-head against doubling and we 
sketch the opposed move: to look explicitly for them in the very core
of mathematics, (differential) geometry. We shall find that, by deforming
differential calculus, the four dimensional coordinate systems
gets the number of degrees of freedom implied by the fermions of
the standard model. So, we propose to use doubling as a tool
to build fermionic actions following classical rules, as 
for instance Einstein-Hilbert.  

\section{A very nice coordinate system is discretized}

Our starting point is, as we have said, four dimensional (1,3) Riemannian
geometry. The core of our proposal is to choose a coordinate system
suitable for scattering experiences; specifically we choose
to use Schwarzchild coordinates, $t, \rho,\theta,\phi$. The volume
form would be then the usual wedging $ dt d\rho d\theta d\phi$ of
anticommuting differentials. The coordinates have the usual
range: $t\in R, \rho \in R^+, \theta \in (0,\pi), \phi \in S^1$. 

Note that this coordination has interesting peculiarities: It is singular
for $\rho=0$, then any calculation must avoid to cross this point. Also,
only two differential terms carry length units. The angles $\theta, \phi$
will get units as an effect of the discretization, but the continuum
limit must be free of units, very much as an asymptotic freedom effect.  
The angle $\theta$ has also a singularity, but it is different from
the one of $\rho$ and we do not  see the most adequate method to
manage it. Perhaps we would to build  a double cover of the space
or to do some trick to define the coordinate  
over $RP^1$.

Having the coordinate system in mind, consider first
 derivatives on any coordinate $x^i$.
\begin{equation}
{\partial A \over \partial x^i}(x) \to 
{ A(\phi_i^+(x,\epsilon)) - A(\phi_i^-(x,\epsilon)) \over \Delta \phi_i }
\end{equation}
The relationship between numerator and denominator,
given by $\Delta \phi_i= \phi_i^+(x,\epsilon)-\phi_i^-(x,\epsilon)$,
could break when the theory is quantized.
Any term containing a derivative of $A$ can thus be 
replaced by a term depending of $A^+,A^-$, and a scale $m_i^{-1}$.

If the coordinate $\rho$ is involved, we have an additional
restriction, as we can not cross $\rho=0$, then we must impose
a positivity condition, $\phi_\rho^\pm(x^\rho) > x^\rho $.

To resume, if we want to calculate and action involving first
derivatives in a four dimensional Riemannian space, we must work
with a duplicated coordinate system, where one of the coordinates
has a restriction to its duplication, and two coordinates must
to be free of length units. Neutrino and quarks come fast to the
mind.

Ideally, we could try to formalize the doubling by rewriting actions
in terms  of (real,complex?) fermions.

%

\section{An action with second derivatives}

Suppose that we have some term in the action needing to evaluate
a second derivative.
The same argument doubling fields above, will then drive us to triple
the number of fields here, as we need to specify three points for
each calculation. But this is not bad: It only
implies that such actions must be discretized by using three "generations"
of coordinate functions.

\begin{equation}
{{A(\phi^r)-A(\phi^c) \over \phi^r-\phi^c} -
    { A(\phi^c)-A(\phi^l) \over \phi^c-\phi^l} 
  \over \phi^+ - \phi^- }
\end{equation} 

Again, the singularity at $\rho=0$ imposes a restriction to the triplication
of the radial component, but is is difficult to trace how this restriction
affects to the whole structure. Mixing and CKM matrix come to mind, if
only as metaphor (But mixing seems to be relevant even without restrictions,
see section 5).

It would be examined what to do when a Lagrangian contains both first
and second derivatives. To tensor both freedoms seems to be the more
reasonable approach, as it clones the known particle spectrum. But
to build the second derivatives by composing the freedom got in the
first ones is also a reasonable assumption, very much as to use a
technicolor technique to get the generations going. 

This latter approach is similar to compose two spin one-half
representations to get one of spin one. But it is interesting to note
that in our case we have an additional discrete ambiguity, namely
to reverse the order of the differences, as its composition counterweights
the change of sign.

Anyway the moral is that, somehow, when trying to build an higher
derivative action, as 
for instance 
gravity, our coordinate
system is forced to approach the shape of the experimentally known fermions.

This technique is not exclusive of gravity; it
could be applied in principle to any higher derivative
model, for instance \cite{sham}. It is interesting to note that
some higher derivative models have been related to antiferromagnetic
fixed points.

Just for enlightenment, lets contemplate one expanded form
of Einstein-Hilbert action

\begin{eqnarray*}
S&=\int (&
-g^{lq} g^{in} {\partial^2 g_{nl} \over \partial x^i \partial x^q}
+g^{lq} g^{in} {\partial^2 g_{ql} \over \partial x^i \partial x^n}
\\&&
-g^{lq} {\partial g^{in} \over \partial x^i} {\partial g_{nl} \over \partial x^q}
+g^{lq} \frac12 {\partial g^{in} \over \partial x^i} {\partial g_{ql} \over \partial x^n}
+g^{lq} \frac12 {\partial g^{in} \over \partial x^l} {\partial g_{ni} \over \partial x^q}
\\&&
-g^{lq} \frac14 g^{in} {\partial g_{ni}  \over \partial x^p} g^{pm} {\partial g_{ml} \over \partial x^q}
-g^{lq} \frac14 g^{in} {\partial g_{ni}  \over \partial x^p} g^{pm} {\partial g_{qm} \over \partial x^l}
+g^{lq} \frac14 g^{in} {\partial g_{ni}  \over \partial x^p} g^{pm} {\partial g_{ql} \over \partial x^m}
\\&&
-g^{lq} \frac14 g^{in} {\partial g_{pn}  \over \partial x^i} g^{pm} {\partial g_{ml} \over \partial x^q}
-g^{lq} \frac14 g^{in} {\partial g_{pn}  \over \partial x^i} g^{pm} {\partial g_{qm} \over \partial x^l}
+g^{lq} \frac14 g^{in} {\partial g_{pn}  \over \partial x^i} g^{pm} {\partial g_{ql} \over \partial x^m}
\\&&
+g^{lq} \frac14 g^{in} {\partial g_{pi}  \over \partial x^n} g^{pm} {\partial g_{ml} \over \partial x^q}
+g^{lq} \frac14 g^{in} {\partial g_{pi}  \over \partial x^n} g^{pm} {\partial g_{qm} \over \partial x^l}
-g^{lq} \frac14 g^{in} {\partial g_{pi}  \over \partial x^n} g^{pm} {\partial g_{ql} \over \partial x^m}
\\&&
+g^{lq} \frac14 g^{in} {\partial g_{nl}  \over \partial x^p} g^{pm} {\partial g_{mi} \over \partial x^q}
+g^{lq} \frac14 g^{in} {\partial g_{nl}  \over \partial x^p} g^{pm} {\partial g_{qm} \over \partial x^i}
-g^{lq} \frac14 g^{in} {\partial g_{nl}  \over \partial x^p} g^{pm} {\partial g_{qi} \over \partial x^m}
\\&&
+g^{lq} \frac14 g^{in} {\partial g_{pn}  \over \partial x^l} g^{pm} {\partial g_{mi} \over \partial x^q}
+g^{lq} \frac14 g^{in} {\partial g_{pn}  \over \partial x^l} g^{pm} {\partial g_{qm} \over \partial x^i}
-g^{lq} \frac14 g^{in} {\partial g_{pn}  \over \partial x^l} g^{pm} {\partial g_{qi} \over \partial x^m}
\\&&
-g^{lq} \frac14 g^{in} {\partial g_{pl}  \over \partial x^n} g^{pm} {\partial g_{mi} \over \partial x^q}
-g^{lq} \frac14 g^{in} {\partial g_{pl}  \over \partial x^n} g^{pm} {\partial g_{qm} \over \partial x^i}
+g^{lq} \frac14 g^{in} {\partial g_{pl}  \over \partial x^n} g^{pm} {\partial g_{qi} \over \partial x^m}
) d\Omega
\end{eqnarray*}
(It is feasible to factorize the metric in a product of vierbeins, and
then to make the gauge explicit, but we will refrain from doing it here)

Note that the number of different terms in the gravity action can
be classified in three kinds, very much as the gauge fields of the
standard model can be ordered inside elements of a algebra of
three components, $C \oplus H \oplus M_3(C)$. Again, this points
to consider NCG (but please do not take too seriously this argument, it
is only a remark out of curiosity).

Either we take the metric or its component vierbein as the 
field to minimize, we confront an action with second derivatives. 
Note also that first derivatives appear always in multiplicative
pairs, which seems difficult to fit with the fermionic action, and that
second derivatives contain crossed terms, which the standard model
could be controlling through coupling constants, instead of masses.

\section{Shadows of NCG, the Higgs and Masses}

Once we have accepted the discretization of derivations, we find us
with a big collection of mass like terms, the mass being related to
the size of our lattices, i.e., to the denominator of each
derivative. 
This result is not nice, as mass must be generated through
couplings with the Higgs.

In some sense, the Higgs should control the discretization. As the mass
of the Higgs goes to infinity, so must go the fermionic
ones. In the high energy, short distances, regime, every mass
is negligible, and we would expect to be performing a fully
non commutative calculus.
{ \it In the low energy, large distances, regime, masses can be taken 
to be infinite, then
its corresponding "distances" approach zero, and we are 
taking usual derivatives}, thus approaching a traditional
commutative action. Also, this effect relates to asymptotic
freedom, as only at low energy our coordinate system needs
to make unobservable the angular variables, while at high
energy they can perform as free particles.
 
Our puzzle lacks of two pieces: one natural geometrical
interpretation of Higgs mass, and a justification for not going
to the infinitesimal limit. Non commutative geometry models
provide the first piece, by relating the Higgs mass with a separation
between multiple sheets of space-time. The above quoted antiferromagnetic
models could be an alternative if we want to remain in the statistical
world of lattice QFT.

To see how NCG is able to hold so complicated interrelations, it is
illustrative to make a one dimensional digression: Take as example
the commutative algebra got from functions in the line through
\beq
f(x) \to F\equiv \pmatrix{f(x+\lambda) & 0 \cr 0 & f(x+\epsilon)}
\eeq
and put $D=\pmatrix{ 0 & 1\over \lambda-\epsilon \cr 1\over \lambda-\epsilon
& 0}$, then $dx\equiv [D,X]= \pmatrix{ 0& -1 \cr 1 & 0} $ and we have
\beq
df\equiv [D,F]= 
\pmatrix{0 & {f(x+\epsilon) - f(x+\lambda) \over \lambda-\epsilon} \cr
  {f(x+\lambda) - f(x+\epsilon) \over \lambda-\epsilon}  & 0}=
 {f(x+\lambda) - f(x+\epsilon) \over \lambda-\epsilon} dx  
\eeq 

The same result can be formulated by using Majid theory of
deformed calculus (\cite{majid}, with $\epsilon=0$), and probably
 also through generalizations
of the tangent grupoid \cite{propio.grupoid}, but NCG is a more general
geometrical theory, and we would prefer to adhere to it. 

As for the second piece, a more fundamental interpretation of QFT is
needed. The recently noticed connection between perturbative
renormalization and Hoft algebras \cite{dirk} could pave the way for a purely
mathematical interpretation of field theory \cite{connes.vietri}.
We could expect to
find us against a deformed differential calculus, its Taylor series
being Feynman ones, and the infinitesimal limit relaxed to a
less exigent one, such as the triviality of
the fully renormalized
theory. 

If we compare the previous example with the actual geometrical
representation of the standard model \cite{connes.cham} we
would suspect that quantization includes the art of keeping
the mass ($1\over \lambda -\epsilon$) finite while the differences are
 driven to zero.

\section{Wondering if Nature learns calculus.}

Lets follow the previous example by adding the generations we suspect to
need to discretize curvatures, second derivatives and so on.

To do it, we triple the previous basis, and expand the Dirac operator
to hold a whole $3 \times 3$ mass matrix $M$, 
\beq
D=\pmatrix{0 & M \cr M^* & 0}, M=M^*
\eeq
and repeat the previous procedure.

Lets put an example using an off-diagonal mass matrix,
\beq 
M=\pmatrix{ 0 & 1/a & 1/b \cr 1/a & 0 & 1/c \cr 1/b & 1/c &0}
\eeq and an initial diagonal matrix
\beq
F\equiv \pmatrix{ f(x+b+c) &&&&& \cr
                  & f(x)   &&&&  \cr
                  && f(x+a+b) &&& \cr
                  &&& f(x+a)  && \cr
                  &&&& f(x+a+b+c) & \cr
                  &&&&& f(x+c)  }
\eeq

Thus,
\beq
dx\equiv [D,X]=
\pmatrix{ & & & 0 & 1 & -1 \cr
          &0& & 1 & 0 & 1 \cr
          & & & -1 & 1 & 0 \cr
         0 & -1 & 1 & & & \cr
        -1 & 0 & -1 & &0& \cr
         1 & -1 & 0 & & & }
\eeq
and (using the notation $[lm...]\equiv f(x+l+m+...)$ and so on)
$$
F' \equiv [D,F] dx^{-1}= 
$$
\beq
\frac12 \pmatrix{
{[abc]-[bc]\over a} + {[bc]-[c]\over b} &
   {[abc]-[bc]\over a} - {[bc]-[c]\over b} &
      {[abc]-[bc]\over a} - {[bc]-[c]\over b} &   \cr
{[a]-[]\over a} - {[c]-[]\over c} &
   {[a]-[]\over a} + {[c]-[]\over c} &
      -{[a]-[]\over a} + {[c]-[]\over c} & 0  \cr
-{[ab]-[a]\over b} + {[abc]-[ab]\over c} &
   -{[ab]-[a]\over b} + {[abc]-[ab]\over c}  &
      {[ab]-[a]\over b} + {[abc]-[ab]\over c}  &   \cr
 \cr  \cr \cr 
& {[a]-[]\over a} + {[ab]-[a]\over b} &
        {[a]-[]\over a} - {[ab]-[a]\over b} &
             {[a]-[]\over a} - {[ab]-[a]\over b} \cr 
0& {[abc]-[bc]\over a}  - {[abc]-[ab]\over c}  &
        {[abc]-[bc]\over a}  + {[abc]-[ab]\over c}  &
           -{[abc]-[bc]\over a}  + {[abc]-[ab]\over c}  \cr
&  - {[bc]-[c]\over b} + {[c]-[]\over c} &
         - {[bc]-[c]\over b} + {[c]-[]\over c} &
             {[bc]-[c]\over b} + {[c]-[]\over c} }
\eeq
Just as in the previous example, if we go to  
the "infinite mass" limit, $a,b,c \to 0$, the matrix $F'$ becomes
simply $f' Id_6$, as we would expect. Observe how entries in $F'$
are always composed of samplings of $f$ in three different points, and how
the off diagonal terms correspond to numerators of second derivatives,
thus going to zero in the limit. 

If we continue the process 
the corresponding $F''$ matrix (defined as $F''\equiv [D,F'] dx^{-1}$)
contains diagonal terms corresponding to second derivatives averaging
two samples of two triplets each (for instance,
$$F''{}_{4,4} =\frac14
( {{f(x+a+b+c)-f(x+a+b)\over c} -{ f(x+a+b)-f(x+a) \over b} \over b} +
     {{f(x+a)-f(x)\over a} -{ f(x+c)-f(x) \over c} \over a} ),
$$ etc.) and off diagonal terms going to zero in the infinite mass limit. Notice
that we have still the freedom to impose "mass relations" in the matrix
$M$ if we want to control, or even nullify, the off diagonal terms. 

Mixing was critical here in order to get a nonzero result, and
it seems so in other examples we have essayed. 
We feel uneasy with this apparent need, as (weak) 
interactions have not been yet introduced at this level.
Anyway, read this toy as an example of the things that 
NCG can formulate.


\section{Conclusions.}

The speculative nature of this article does not let us to derive
solid conclusions. Instead of this, lets to abstract the sugerences
raised along the note:

\begin{itemize}
\item The existence of four fundamental particles $e,\nu,u,d$ is related
to the construction of a coordinate system adequate to Minkowskian
space, perhaps Schwarzchild coordinates  $t,\rho,\theta,\phi$.
\item Doubling (quirality) of fermions relates to the ambiguity to choose
the two points needed to build regularized derivatives.
\item Undoubling of neutrino could come from the restriction on the
discretization of the radial coordinate.
\item Asymptotic freedom (confinement?) is due to the absence of length
scale in two of the coordinates, say $\theta,\phi$
\item Tripling of generations comes from the ambiguity to choose the three
points needed to build regularized second derivatives. 
\item Mixing (weak CP etc) of generations could relate also to the
restriction on
the discretization of second partial 
derivatives involving $\rho$ (and $\theta?$)
\item The Higgs (alternatively space sheets, alternatively antiferromagnetic
vacuum) controls the scales which are applied in each derivation. 
\item Triviality is equivalent to ask for a good limit
$\Delta x^i \to 0$ in every
derivative. 
\item Gauge fields will appear if we try to apply the regularization procedure
to a general covariant action. 
\end{itemize}

Obviously the risky bet here is the relationship between coordinates
and fermions. Once one gets to live with it, each conjecture can try
to prove its validity in an independent form. But its combination 
drives us to look for a strange form of induced gravity: the limit
$M_{Higgs} \to \infty$ of the Standard Model should be equivalent to
some gravity-like action, and the fermionic matter field would be
pumped to existence in its quantum version, as a lateral effect of
the renormalization process.

We would like to note that quantum theory is by itself a regularization
not of geometry but of variational calculus, as Feynman integral
teach us.
The nature of the recipe relating deformation and quantization
has not been not examined yet (but search \cite{propio.path} for clues),
we would expect Plank constant to emerge as the parameter relating
both techniques.

\section{Acknowledgements}

This work was first imagined while contemplating the beautiful Salerno 
bay at Vietri, and then doubted one time and another, while
coalescing it along the way to return home.
Hospitality of the organizing team of Vietri school, and
specially of Julio Guerrero is cheerfully acknowledged.

Financial travel support from project MECxxx.yyy, 
resources and comments from DFTUZ, as usual, and the opportunity to steal
some spare time from my work at Telefonica I+D, have been really welcome.

\end{document}